# Bulk OsO$_2$ Single Crystals: Superior Catalysts for Water Oxidation

*Guojian Zhao[#], Zhihao Li[#], Ziang Meng\*, Shucheng Wang, Li Liu, Zhiyuan Duan, Xiaoning Wang, Hongyu Chen, Yuzhou He, Jingyu Li, Sixu Jiang, Xiaoyang Tan, Qinghua Zhang\*, Qianfan Zhang\*, Peixin Qin\*, Zhiqi Liu\**

G. Zhao, Z. Li, Z. Meng, S. Wang, L. Liu, Z. Duan, H. Chen, J. Li, S. Jiang, X. Tan, Q.F. Zhang, P. Qin, Z. Liu

School of Materials Science and Engineering, Beihang University, Beijing 100191, China

E-mail: mengza@buaa.edu.cn; qianfan@buaa.edu.cn; qinpeixin@buaa.edu.cn; zhiqi@buaa.edu.cn

Y. He, Q.H. Zhang

Beijing National Laboratory for Condensed Matter Physics, Institute of Physics, Chinese Academy of Sciences, Beijing 100190, China.

E-mail: zqh@iphy.ac.cn

G. Zhao, Z. Meng, L. Liu, Z. Duan, H. Chen, J. Li, S. Jiang, X. Tan, P. Qin, Z. Liu

State Key Laboratory of Tropic Ocean Engineering Materials and Materials Evaluation, Beihang University, Beijing 100191, China.

X. Wang

The Analysis & Testing Center, Beihang University, Beijing, 100191, China

[#]These authors contributed equally to this work.

**Keywords**: OsO$_2$ single crystals, OER performance, RuO$_2$, high stability, bulk single crystal catalysts.

**Abstract**: Although rutile RuO$_2$ has been a well-known and almost the best oxygen evolution reaction (OER) catalyst, the OER properties for the similar rutile oxide OsO$_2$ with the same group element with Ru have been unknown, mainly due to long-standing synthesis difficulties. In this work, we report the successful synthesis of high-quality OsO$_2$ single crystals, and the ground micrometer-size single crystals are chemically stable in alkaline solutions and exhibit robust OER performance. In sharp contrast, OsO$_2$ nanopowder reacts quickly with KOH solutions and cannot work for OER. Compared with commercial RuO$_2$ nanopowder, the OsO$_2$ single crystals show comparable catalytic current densities, remarkably lower overpotentials at high current densities and better stability. These findings question the universal applicability of nanoscaling and highlight crystal integrity as a key descriptor for achieving stable and efficient OER electrocatalysis.

# 1. Introduction

The transition to a sustainable energy future depends on the efficient utilization of renewable energy sources, such as solar and wind, necessitating advanced energy conversion and storage technologies.[1-3] The oxygen evolution reaction (OER) is a critical process in electrochemical energy conversion and storage systems, underpinning technologies such as water electrolysis, metal-air batteries, and renewable energy systems.[4-8] This critical four-electron-process involves the oxidation of water molecules to produce oxygen gas, requiring efficient catalysts to overcome its sluggish kinetics and high overpotential.[9-12] Various catalysts, incorporating diverse catalytic centers, spin states, and surface modifications, have been developed to enhance OER efficiency and reduce energy consumption in electrochemical processes.[13-22] However, achieving a balance between performance, cost, and stability for industrial-scale water electrolysis remains a significant challenge, as no catalyst has yet fully met these criteria. The development of robust and well-defined OER electrocatalysts remains a critical research priority for advancing fundamental understanding of electrochemical water oxidation.

Among precious metal-based catalysts, ruthenium dioxide ($RuO_2$) is widely regarded as a benchmark electrocatalyst for OER due to its exceptional catalytic performance in both acidic and alkaline electrolytes.[23-30] The efficacy of $RuO_2$ derives from its stable rutile crystal structure, which provides robust active sites and a favorable electronic configuration for the formation of oxygen intermediates. Iridium dioxide ($IrO_2$), which shares the same rutile structure and is positioned diagonally to ruthenium in the periodic table, also exhibits comparable catalytic performance, as confirmed by numerous experimental studies.[31-33]

In sharp contrast, osmium dioxide ($OsO_2$), a compound from the same group as ruthenium, has received rather limited attention for OER experimental investigations. Despite theoretical studies highlighting similarities in the electronic structures of $OsO_2$ and $RuO_2$,[34-37] to our best knowledge, there has been no experimental report on the OER properties of $OsO_2$. The reasons could be understood from the different material forms - single crystals and polycrystalline powder: (1) $OsO_2$ single crystals are not commercially available and are greatly challenging to be synthesized owing to the possible generation of volatile and highly toxic intermediates. Specifically, the precursor material, metallic osmium (Os) powder, is unstable at room temperature and can slowly be oxidized to yield volatile osmium tetroxide ($OsO_4$), which is highly toxic. Consequently, up to now, there have been only two experimental reports on the synthesis of $OsO_2$ single crystals: one in 1969[38] and the other in 2004.[39] Although $OsO_2$ single crystals have been reported previously for electrical transport and structural studies, their electrocatalytic properties have not been experimentally explored. (2) Polycrystalline $OsO_2$ powder is commercially available and it typically contains nano-size (10-100 nm) particles. However, we found that the commercial $OsO_2$ nanopowder quickly reacts with alkaline KOH solutions and is not stable for standard OER tests. Thus, the intrinsic OER properties of $OsO_2$ remain mysterious and unknown. In this context, $OsO_2$ serves as a model system that enables experimental access to the intrinsic OER behavior of a rutile-type oxide in a structurally well-defined form.

In this study, we report the successful synthesis of high-quality bulk $OsO_2$ single crystals. Remarkably, micron-sized $OsO_2$ single crystals demonstrate superior OER activity compared to $RuO_2$ at higher overpotentials. The larger particle size and single-crystal nature of our $OsO_2$ significantly enhance its stability during electrochemical processes, enabling stable operation over 120 h at a current density of 10 mA·cm$^{-2}$. In contrast, commercially available nanoscale $OsO_2$ powders dissolve within seconds under identical OER testing conditions. Density Functional Theory calculations were performed to assess the reactivity of various $OsO_2$ surfaces for the OER. The (110) surface exhibited the highest OER activity where hydroxyl (OH) binding was the rate-limiting step. Its octahedrally coordinated osmium (Os) active sites enable minimal electron transfer during *OH formation, promoting efficient desorption and subsequent intermediate formation. Our results indicate that an overemphasis on maximizing specific surface area through particle size reduction may adversely affect catalyst stability, emphasizing the critical need to optimize particle size to achieve an effective balance between chemical stability and catalytic activity. This study highlights the potential of bulk single-crystal catalysts and identifies crystal integrity as an important descriptor for balancing activity and stability in electrocatalyst design.

## 2. Results and discussion

**Synthesis and characterization of $OsO_2$ single crystals**

$OsO_2$ single crystals were synthesized via a general two-step approach including chemical vapor transport (Figure 1a). Figure 1b presents an optical image of the synthesized $OsO_2$ single crystal, exhibiting characteristic dimensions of approximately 1 mm. The lustrous, golden surface likely corresponds to low-index crystallographic planes, which are typically characterized by lower surface energies due to reduced dangling bonds and increased coordination numbers. This observation underscores the high quality of the synthesized single crystal. Structurally, $OsO_2$ adopts the rutile crystal structure, analogous to $RuO_2$ (Figure 1c), which has been extensively investigated in recent literature. Figure 1d illustrates the temperature-dependent resistivity of the $OsO_2$ single crystal, measured from 2 K to 300 K. At room temperature, the resistivity of $OsO_2$ is approximately 22.4 μΩ·cm, notably lower than that of bulk $RuO_2$ single crystals (~35.2 μΩ·cm).[40] This enhanced conductivity in $OsO_2$ establishes a critical foundation for its application in electrochemical catalytic water oxidation. Rietveld refinement analysis confirms that $OsO_2$ possesses a tetragonal rutile structure (space group $P42/mnm$), with lattice parameters determined as $a = b = 4.4971$ Å and $c = 3.1841$ Å (Figure 1e). The phase identity of the crystal is further supported by Raman spectroscopy (Figure S2a), which reveals three characteristic vibrational modes ($E_g$, $A_{1g}$, and $B_{2g}$) that are consistent with the rutile-type $OsO_2$ structure. These values are in close agreement with previously reported lattice constants[39] and theoretical calculation for $OsO_2$(Figure S2b), indicating robust structural integrity and minimal defects introduced during the crystal synthesis and grinding stages.

Transmission electron microscopy (TEM) characterizations provide detailed insights into the sample's structure. The selected area electron diffraction pattern in Figure 2a reveals the $(01\bar{1})$ and $(1\bar{1}0)$ planes along the [111] zone axis, confirming the crystalline orientation. High-resolution TEM imaging (Figure 2b) and magnified views (Figure 2c) highlight the atomic arrangement with a scale bar of 5 nm and 1 nm, respectively. Electron energy loss spectroscopy (Figure 2d) identifies the O $K$-edge and Os $M$-edge, indicating the elemental composition. High-angle annular dark-field (HAADF) imaging (Figure 2e) and elemental mapping for Os and O, along with the combined Os&O map (Figure 2f-h), further demonstrate the uniform distribution and high quality of the as-grown single crystal, consistent with the rutile phase.

**OER performance of $OsO_2$ single crystals and nanopowder**

Figure 3a displays the linear sweep voltammetry (LSV) curves for two $OsO_2$ samples assessed in a 1 mol·L$^{-1}$ KOH electrolyte. The LSV curve for $OsO_2$ single crystals exhibits a pronounced increase in current density, surpassing 75 mA·cm$^{-2}$ at potentials above 1.8 V vs. RHE, indicative of markedly enhanced OER activity compared with the nanopowder sample. This pronounced enhancement suggests efficient charge transfer and the presence of highly active catalytic sites, likely attributable to the well-defined low-index crystallographic planes (*e.g.*, {110}), which are commonly observed in rutile-type single-crystal structures. In contrast, the LSV curve for $OsO_2$ nanopowder demonstrates negligible OER catalytic activity, potentially due to structural instability, high surface defect density, or diminished electronic conductivity associated with its nanoparticulate morphology. Notably, two distinct peaks are observed in the nanopowder curve at ~1.35 V and 2.0 V vs. RHE. These peaks are hypothesized to reflect electrochemical surface oxidation processes or the formation of intermediate species (OsOOH or higher oxidation states of Os), rather than direct contributions to OER catalysis.

The Figure 3b-e show the distinct surface characteristics of the prepared working electrodes. The ground $OsO_2$ single crystals, with particle sizes ranging from approximately 1 to 100 μm, exhibit a lustrous gold coloration consistent with the bulk material (Figure 3b and 3c). In contrast, the commercial $OsO_2$ nanopowder, with significantly smaller particle sizes of approximately 10 to 100 nm, displays a dark black appearance (Figure 3d and 3e). The nanoscale dimensions of the commercial powder markedly enhance its surface reactivity, which detrimentally impacts the stability of OER catalysis. Notably, the cyclic voltammetry (CV) curves of commercial $OsO_2$ nanopowder and single crystals (Figure S3a) reveal a significant difference in electrocatalytic behavior. A closer inspection of the nanopowder CV curves (Figure S3b) shows a gradually diminishing redox peak over repeated cycles, indicating poor electrocatalytic stability likely caused by surface dissolution at the nanoscale.

The X-ray Photoelectron Spectroscopy (XPS) was employed to probe the surface chemical states of the ground $OsO_2$ single crystals and commercial $OsO_2$ nanopowder. As shown in the Figure 3f and 3g, the XPS spectrum of commercial $OsO_2$ nanopowder and ground $OsO_2$ share broadly similar core-level features. The high-resolution Os 4$f$ spectra for both samples exhibit distinct peaks, and the corresponding O 1$s$ spectra reveal a significant proportion of lattice oxygen. On the contrary, the commercial $OsO_2$ nanopowder displays a markedly reduced lattice oxygen content in its O 1$s$ spectrum, accompanied by notably weaker Os 4$f$ peak intensities. These surface characteristics are consistent with the rapid electrochemical deactivation observed for the $OsO_2$ nanopowder in alkaline electrolyte. These findings suggest that the nanoscale and polycrystalline nature of the commercial nanopowder leads to inferior material quality, characterized by abundant surface defects and oxygen vacancies. Such defects enhance the instability of the nanopowder and thus compromise its stability during OER catalysis. Conversely, the ground $OsO_2$ single crystals retain a largely intact lattice oxygen framework with fewer grain boundaries and defects, contributing to their enhanced chemical stability and sustained OER performance.

**Superior catalytic performance of $OsO_2$ single crystals compared to $RuO_2$**

To further evaluate the performance of $OsO_2$, commercial $RuO_2$ nanopowder was used as a benchmark. LSV was applied to compare the OER catalytic activity of ground $OsO_2$ single crystal and commercial $RuO_2$ in $O_2$ saturated 1 mol·L$^{-1}$ KOH. As depicted in Figure 4a, commercial RuO$_2$ nanopowder exhibits a lower onset overpotential compared to OsO$_2$. Intriguingly, at higher potentials, the current density of OsO$_2$ increases more rapidly than that of RuO$_2$, indicating superior high-current-density OER performance. Notably, OsO$_2$ surpasses RuO$_2$ above 1.66 V vs. RHE, achieving a maximum current density of approximately 119.8 mA·cm$^{-2}$ at 2.0 V vs. RHE, compared to only 79.9 mA·cm$^{-2}$ for RuO$_2$ at the same potential. This elevated current density suggests a faster kinetics of charge transfer during the OER process and the higher slope of LSV curve under high overpotential may indicate intrinsically favorable oxygen evolution kinetics at high overpotentials. The overpotentials at four catalytic current density values of 10, 25, 50, and 100 mA·cm$^{-2}$ were compared in Figure 4b. At 10 mA·cm$^{-2}$, RuO$_2$ requires a lower overpotential of 360 mV compared to 390 mV for OsO$_2$. However, at higher current densities of 25, 50, and 100 mA·cm$^{-2}$, OsO$_2$ exhibits lower overpotentials (451, 546, and 723 mV, respectively) than RuO$_2$ (457, 592, and 905 mV, respectively). Furthermore, the Tafel slopes, shown in Figure 4c, are 43 mV·dec$^{-1}$ for OsO$_2$ and 76 mV·dec$^{-1}$ for RuO$_2$, providing additional evidence of OsO$_2$'s superior OER kinetics. It is worth noting that the particle size of commercial RuO$_2$ nanopowder is relatively small and comparable to that of commercial OsO$_2$ nanopowder (Figure S4), suggesting a similarly high density of catalytic sites. Nevertheless, OsO$_2$ single crystals still exhibit superior high-current catalytic performance.

To provide further insight into the intrinsic catalytic activity beyond geometric current density, additional normalization analyses based on the electrochemically active surface area (ECSA), roughness factor (RF), and turnover frequency (TOF) were carried out. Electrodes were prepared using the same drop-casting procedure as described in the Methods, and the independently prepared electrodes exhibited highly reproducible OER polarization curves under identical electrochemical conditions (Figure S5).

The ECSA was estimated from double-layer capacitance measurements conducted in a non-faradaic potential region (Figure S6). The obtained ECSA values for OsO$_2$ single crystals and commercial RuO$_2$ nanopowder are 5.03 and 19.6 cm$^2$, respectively, indicating a substantially lower density of electrochemically accessible active sites for OsO$_2$. After normalizing the current density by ECSA, the resulting $j_{ECSA}$ curves (Figure S7a) show that OsO$_2$ single crystals and RuO$_2$ nanopowder exhibit comparable intrinsic activity at low potentials. Notably, as the applied potential increases, the $j_{ECSA}$ of OsO$_2$ single crystals progressively and significantly exceeds that of RuO$_2$ (Figure S7b), demonstrating a superior intrinsic catalytic activity of OsO$_2$ in the high-current regime. Consistently, the roughness factors derived from the ratio of ECSA to geometric area are approximately 71 for $OsO_2$ single crystals and 277 for $RuO_2$ nanopowder, reflecting their markedly different particle sizes and surface morphologies. Furthermore, a conservative lower-bound estimation of the turnover frequency was performed by assuming all metal atoms to be catalytically active (Figure S8). The potential-dependent TOF analysis independently supports an intrinsically higher reaction rate of $OsO_2$ single crystals at elevated overpotentials.

A chronopotentiometry test was conducted to evaluate the long-term stability of $OsO_2$ single-crystal catalysts in the same 1mol·L$^{-1}$ KOH solution under a benchmark current density of 10 mA·cm$^{-2}$. As shown in Figure 4d, no obvious voltage degradation was observed during continuous operation for over 120 h. In contrast, $RuO_2$ nanopowder exhibits a distinct inactivation at ~10 h and a significant voltage degradation at ~35 h. In addition, inductively coupled plasma mass spectrometry (ICP-MS) analysis of the electrolyte collected during the stability test shows only a low level of osmium dissolution, which tends to saturate over time (Figure S9). Compared with the commercial $OsO_2$ nanopowder, the bulk single crystal form of the obtained $OsO_2$ with micron size greatly enhanced its stability in alkaline environments, while maintaining a satisfactory catalytic activity. This improved durability is consistent with the intrinsic characteristics of single-crystal materials, including a reduced density of high-energy grain boundaries, which mitigates corrosion and surface reconstruction under OER conditions. These results further underscore the importance of crystal integrity in suppressing catalyst degradation during alkaline OER operation.

To further elucidate the structural and chemical stability of $OsO_2$ single crystals under OER conditions, a series of post-OER characterizations were performed after prolonged electrochemical operation. Raman spectroscopy and X-ray photoelectron spectroscopy were conducted after 10 h of continuous OER at a current density of 10 mA·cm$^{−2}$. As shown in Figure S10, the Raman spectra after OER preserve the characteristic vibrational modes of rutile $OsO_2$ ($E_g$, $A_{1g}$, $B_{2g}$), with no additional peaks or noticeable peak broadening observed, indicating the retention of the rutile phase during OER operation. Consistently, XPS analysis (Figure S11) shows that the Os 4$f$ core-level spectra before and after OER can be fitted with a single spin-orbit doublet corresponding to $Os^{4+}$, with negligible changes in binding energy or peak width. Although minor variations in the relative contributions of surface-associated oxygen species are observed in the O 1$s$ spectra after OER, their binding energies remain essentially unchanged, suggesting surface reorganization rather than bulk oxidation or phase transformation. To assess the long-term structural integrity of $OsO_2$ single crystals, transmission electron microscopy and selected-area electron diffraction analyses performed after 120 h of continuous OER operation reveal sharp diffraction spots characteristic of rutile $OsO_2$ (Figure S12a), together with HAADF-STEM images that directly resolve the ordered arrangement of Os atomic columns (Figure S12b), with no evidence of amorphization or secondary phase formation. Taken together, these results indicate that $OsO_2$ single crystals maintain both structural integrity and chemical robustness under the investigated alkaline OER conditions.

One of the major challenges for the OER applications for $RuO_2$ is its poor chemical stability, for example, under the 0.5 M $H_2SO_4$ solution, the stable OER activity can only exists for less than 10 h.[41,42] Under the 1 M alkaline KOH solution, as tested in our experiments, $RuO_2$ powder is stable within the first 10.7 h and the deactivation shows up subsequently. In sharp contrast, our $OsO_2$ single crystals are chemically stable for over 120 h. This merit could well overcome the poor chemical stability of $RuO_2$ for OER applications.

Density functional theory (DFT) calculations were performed to investigate the catalytic activity of various $OsO_2$ surfaces for the OER. Five low-index surfaces, (100), (001), (110), (101), and (111), were selected for analysis due to their experimentally confirmed stability. The Adsorption Evolution Mechanism (AEM) was employed to model the OER under alkaline conditions,[43,44] with Os atoms designated as the active sites. Figure 5a presents the Gibbs free energy profiles for the OER on these surfaces. Notably, the $OsO_2$ (110) surface (Figure 5c) exhibited the highest OER activity, with a theoretical overpotential of 0.31 V. The potential-determining step was identified as the binding of *OH, with an energy barrier of 1.54 eV. The formation of *O and *OOH intermediates displayed comparable energy barriers, while the formation of $O_2$ molecules was found to be energetically favorable. In contrast, other $OsO_2$ surfaces (Figure 5a and 5b) showed significantly exothermic *OH formation but required higher energy for the formation of *OOH and $O_2$ molecules, indicating less favorable OER kinetics.

The coordination environment and charge transfer dynamics at the active sites of various $OsO_2$ surfaces were analyzed to elucidate the superior catalytic activity of the (110) surface (Figure 5d). Charge density difference calculations reveal significant electron exchange between Os atoms and *OH across all surfaces. However, the (110) surface, characterized by octahedrally coordinated Os active sites consistent with the bulk phase, exhibits the lowest degree of electron transfer during *OH formation. This suggests a moderately weaker *OH binding affinity, which facilitates subsequent desorption and the formation of further intermediates. In contrast, the incomplete coordination of Os atoms on other surfaces results in stronger hydroxyl adsorption and pronounced electron transfer, elevating the energy barrier for *OOH formation.

Consequently, the OsO$_2$ (110) surface is optimal for balanced *OH adsorption, featuring fully coordinated Os atoms surrounded by oxygen atoms. These distinctive active sites underpin the enhanced catalytic performance of the (110) surface in the OER. Given that the (110) surface exhibits one of the lowest surface energies, comparable to that of the (001) surface (Figure S13), it is thermodynamically favorable and therefore likely to be exposed in bulk OsO$_2$ single crystals, making it relevant to the OER process.

It is worth noticing that Os is an expensive element. But a remarkable point on our newly synthesized OsO$_2$ single crystals is that is that they are superiorly stable chemically (over 120 h) under the alkaline environments for the OER. That means they are basically intact after the water splitting. Hence, there is no nominal loss or consumption of these experimental materials during the OER. Therefore, this shall not be a major drawback for fundamental research or practical applications.

## 3. Conclusion

To briefly conclude, motivated by the curiosity on the intrinsic OER performance of OsO$_2$, which is a sister material of RuO$_2$ but its OER properties have never been unveiled, we have successfully grown high-quality OsO$_2$ single crystals. The ground OsO$_2$ single crystals demonstrate outstanding durability and high-current OER performance, outperforming commercial RuO$_2$ nanopowder. It was found that, despite surface defects introduced by grinding, the bulk-derived crystalline integrity of OsO$_2$ single crystals with a lower defect density than commercial nanopowders plays a key role in stabilizing the catalyst during OER. DFT calculations further reveal that the (110) crystallographic surface, with optimally coordinated Os active sites, could largely promote the efficient OER kinetics through balanced hydroxyl adsorption. This work highlights bulk single-crystal catalysts as a powerful platform for probing intrinsic electrocatalytic behavior.

**Methods**

*Sample preparation*

OsO$_2$ single crystals were synthesized using a two-step method (Fig. 1a). In the first step, a stoichiometric mixture of Os powder (3 g, 99.8% purity) and NaBrO$_3$ powder (1.57 g, 99.5% purity)—determined by the molar ratio of Os to O in OsO$_2$—was homogenized and vacuum-sealed in a quartz tube (base pressure: 10$^{-5}$ Torr). The sealed tube was heated in a tube furnace at 300 °C for 48 hours (preheating) before being ramped to 650 °C (heating rate: 1 °C/min) for an additional 48-hour reaction. The furnace was then cooled to room temperature (1 °C/min), yielding polycrystalline $OsO_2$ powder at one end of the tube. (For larger Os precursor masses, an alternative stepwise heating protocol was developed to improve operational safety during the first-stage oxidation process. In this protocol, the temperature is increased in multiple stages with low ramping rates and intermediate dwelling steps below the onset of $NaBrO_3$ decomposition, allowing gradual oxygen release under evacuated sealed-tube conditions (Fig. S1). Using this protocol, reactions involving up to (4.0 + 0.1) g Os powder (with an additional 0.1 g as a protective excess) and 2.11 g $NaBrO_3$ were carried out reproducibly. This alternative protocol was not used for the experiments presented in the main text and is provided as a safer option for larger precursor loads.)

In the second step, the tube was transferred to a dual-zone furnace, with the $OsO_2$-loaded end maintained at 950 °C and the opposite end at 890 °C for 14 days to facilitate chemical vapor transport growth of single crystals. Heating rates were 50 °C/h below 800 °C and 10 °C/h above 800 °C; cooling to room temperature was uniformly conducted at 1 °C/min.

After crystal growth, the $OsO_2$ single crystals were mechanically separated from the quartz tube and cleaned by ultrasonication in deionized water for 10 min to remove residual transport agents. Under the standard synthesis conditions described above, starting from 3 g of Os precursor, a typical batch yields on the order of ~2 g of crystalline $OsO_2$ products, based on the mass of collectable single crystals. It should be noted that a small fraction of very fine, powder-like crystals formed during growth is difficult to collect and therefore is not included in this estimate. Commercial $OsO_2$ and $RuO_2$ powders were obtained from Aladdin. Prior to electrochemical measurements, the as-grown $OsO_2$ single crystals were manually ground using an agate mortar and pestle to obtain micrometer-sized particles. Throughout this work, the term "ground $OsO_2$ single crystals" refers to these manually ground, bulk-derived micrometer-sized particles. No mechanical milling or high-energy grinding was involved.

*Electrochemical measurements*

The electrochemical tests were carried out in $O_2$ saturated 1 mol·L$^{-1}$ KOH solution on the CHI660E electrochemical workstation. Ag/AgCl filled with saturated KCl solution and graphite rod were used as reference and the counter electrodes. The potential value was calibrated with respect to the reversible hydrogen electrode (RHE), which in 1 mol·L$^{-1}$ KOH the formula is $E_{RHE} = E_{Ag/AgCl} + 1.023$ V. To prepare the working electrode, the obtained bulk $OsO_2$ single crystals were manually ground into powder in an agate mortar and a glassy carbon electrode with a diameter of 3 mm was used as the working electrode. Typically, 5 mg catalyst were suspended in 1 mL isopropanol-water solution with a volume ratio of 1:1 with 50 μL Nafion solution to form a homogeneous ink assisted by ultrasound for 1 h. Then, 10 μL of the ink was spread onto the surface of the glassy carbon electrode (mass loading: 0.71 mg·cm$^{-2}$) by a micropipette and dried under room temperature.

Cyclic voltammetry curves were measured at a scan rate of 50 mV·s$^{-1}$. Linear sweeping voltammograms were obtained at a scan rate of 10 mV·s$^{-1}$. The magnetic stirring was operating during the whole measurement process which enables the generated oxygen bubbles to detach from the surface in time to prevent their impact on the effective active area. The LSV curve was replotted as potential vs log current density to obtain Tafel plots for quantification OER performance. Chronopotentiometry measurement method was adopted to study the stability of catalysts, with a constant current density of 10 mA·cm$^{-2}$.

It should be noted that the electrochemical conditions employed in this study (room temperature, aqueous alkaline electrolyte) differ fundamentally from the strongly oxidative and high-temperature conditions under which $OsO_4$ formation is typically observed.

*Sample characterization*

X-ray diffraction measurements were performed via a Bruker D8 Advance diffractometer with Cu-Kα radiation ($\lambda$ = 1.54184 Å). Transmission electron microscopy characterization was conducted by a JEOL NeoArm system operated at 200 kV. X-ray photoelectron spectroscopy was collected by a Thermo Fisher XI+ system. Raman spectroscopy measurements were carried out by a HORIBA LabRAM Odyssey spectrometer. Inductively coupled plasma mass spectrometry (ICP-MS) measurements were performed using a NexION 5000 multi-quadrupole ICP-MS system. For electrical transport measurements, electrical contacts were constructed by an Al wire bonder and the measurements were implemented by a physical properties measurement system.

*Density-functional theory calculations*

All the density-functional theory (DFT) calculations were performed using the Vienna Ab initio Simulation Package (VASP).[45] The Perdew-Burke-Ernzerhof (PBE) functional under the generalized gradient approximation (GGA) was used to describe the exchange–correlation potential.[46] The electron-ion interactions were treated by the projector augmented-wave (PAW) approximation.[47] A cutoff energy of 500 eV was employed for the plane-wave basis set of all calculations. The convergence criteria of structure optimization were chosen with the maximum force on each atom less than 0.01 eV/Å, with an energy change of less than 1 × 10$^{-6}$ eV. The Monkhorst-Pack k-point scheme[48] was used for sampling the Brillouin zones, with a fine resolution of $(0.03 \times 2\pi)$ Å$^{-1}$ All the surfaces were constructed using a triple-layered $(3 \times 3)$ periodic slab, with an Os/O ratio of 1:2. In order to avoid the interaction effect of periodicity, a vacuum layer of 20 Å was introduced along the $z$ axis. The Grimme's DFT-D3 semi-empirical correction was used to describe the vdW interaction.[49] The second and third order force constants were calculated by VASP, Phonopy[50] and Phono3py,[51,52] with the finite displacement method based on $3 \times 3 \times 3$ supercells (162 atoms) and $2 \times 2 \times 2$ supercells (48 atoms), respectively. The macroscopic dielectric constant tensor for both pristine and displaced (along the eigenvector of Raman active modes) primitive cells (6 atom) were calculated with density functional perturbation theory (DFPT). From the force constants, Raman spectra was simulated using Phonopy-Spectroscopy.[53]

The elementary calculations were considered according to the following scheme.

$$OH^- + * \rightarrow *OH + e^-$$
$$OH^- + *OH \rightarrow *O + H_2O + e^-$$
$$OH^- + *O \rightarrow *OOH + e^-$$
$$OH^- + *OOH \rightarrow O_2 + H_2O + e^-$$

To assess the free energy change of each step, first, we calculated the electronic energy change ($\Delta E_{ads}$) of key intermediates.

$$\Delta E_{OH} = E(*OH) + 0.5 \times E(H_2) - E(H_2O) - E(*)$$
$$\Delta E_O = E(*O) + E(H_2) - E(H_2O) - E(*)$$
$$\Delta E_{OOH} = E(*OOH) + 1.5 \times E(H_2) - 2 \times E(H_2O) - E(*)$$

The $\Delta G_{ads}$ values are calculated by[54]

$$\Delta G_{ads} = \Delta E_{ads} + \Delta E_{ZPE} + \Delta U + \Delta(PV) - T\Delta S$$

where $\Delta E_{ZPE}$, $\Delta S$ and $\Delta U + \Delta(PV)$ denote the zero-point energy correction, entropy correction and enthalpy correction, respectively.

The Gibbs free energy change of four steps is calculated as

$$\Delta G_1 = \Delta G_{OH} - eU$$
$$\Delta G_2 = \Delta G_O - \Delta G_{OH} - eU$$
$$\Delta G_3 = \Delta G_{OOH} - \Delta G_O - eU$$
$$\Delta G_4 = 4.92 \text{ eV} - \Delta G_{OOH} - eU$$

where $U$ is the potential under standard conditions versus RHE.

The step with the maximum Gibbs free energy change $\Delta G_{max}$ is identified as the rate-determining step (RDS).

$$\Delta G_{max} = \max(\Delta G_1, \Delta G_2, \Delta G_3, \Delta G_4)$$

The theoretical OER overpotential can be determined by

$$\eta = \Delta G_{max}/e - 1.23 \text{ V}$$

The surface energies of the low-index facets are calculated by

$$E_{surface} = (E_{slab} - E_{bulk})/2A$$

Where $E_{slab}$ is the total energy of each (3 × 3) slab, $E_{bulk}$ is the total energy of the infinite system per (3 × 3) supercell, and A is the area of each slab. The (1/2) factor takes into account the existence of two free surfaces for each slab.[55]

**Acknowledgements**

P.Q. acknowledges financial support from the National Natural Science Foundation of China (no. 52401300). Z.M. acknowledges financial support from the National Natural Science Foundation of China (no. 524B2003). L.L. acknowledges financial support from the National Natural Science Foundation of China (no. 525B2008). Zhiqi L. acknowledges financial support from the National Natural Science Foundation of China (nos. 52425106, 52121001 and 52271235). Zhiqi L. acknowledges financial support from the National Key R&D Program of China (nos. 2022YFB3506000 and 2022YFA1602700). Zhiqi L. acknowledges financial support from the Beijing Natural Science Foundation (no. JQ23005). Q.Z. acknowledges financial supports from the National Natural Science Foundation of China (no. 52322212) and the National Key R&D Program of China (nos. 2023YFA1406300 and 2024YFA1409500). P.Q. acknowledges funding from the China National Postdoctoral Program for Innovative Talents (no. BX20230451) and from the China Postdoctoral Science Foundation (no. 2024M754058). This work is supported by National Natural Science Foundation of China (no. U25A20244). This work is supported by the Fundamental Research Funds for the Central Universities. The authors acknowledge the Analysis & Testing Center of Beihang University for the assistance.

**Supporting Information**

Supporting Information is available from the Wiley Online Library or from the author.

**Data Availability Statement**

The data that support the findings of this study are available from the corresponding author upon reasonable request.

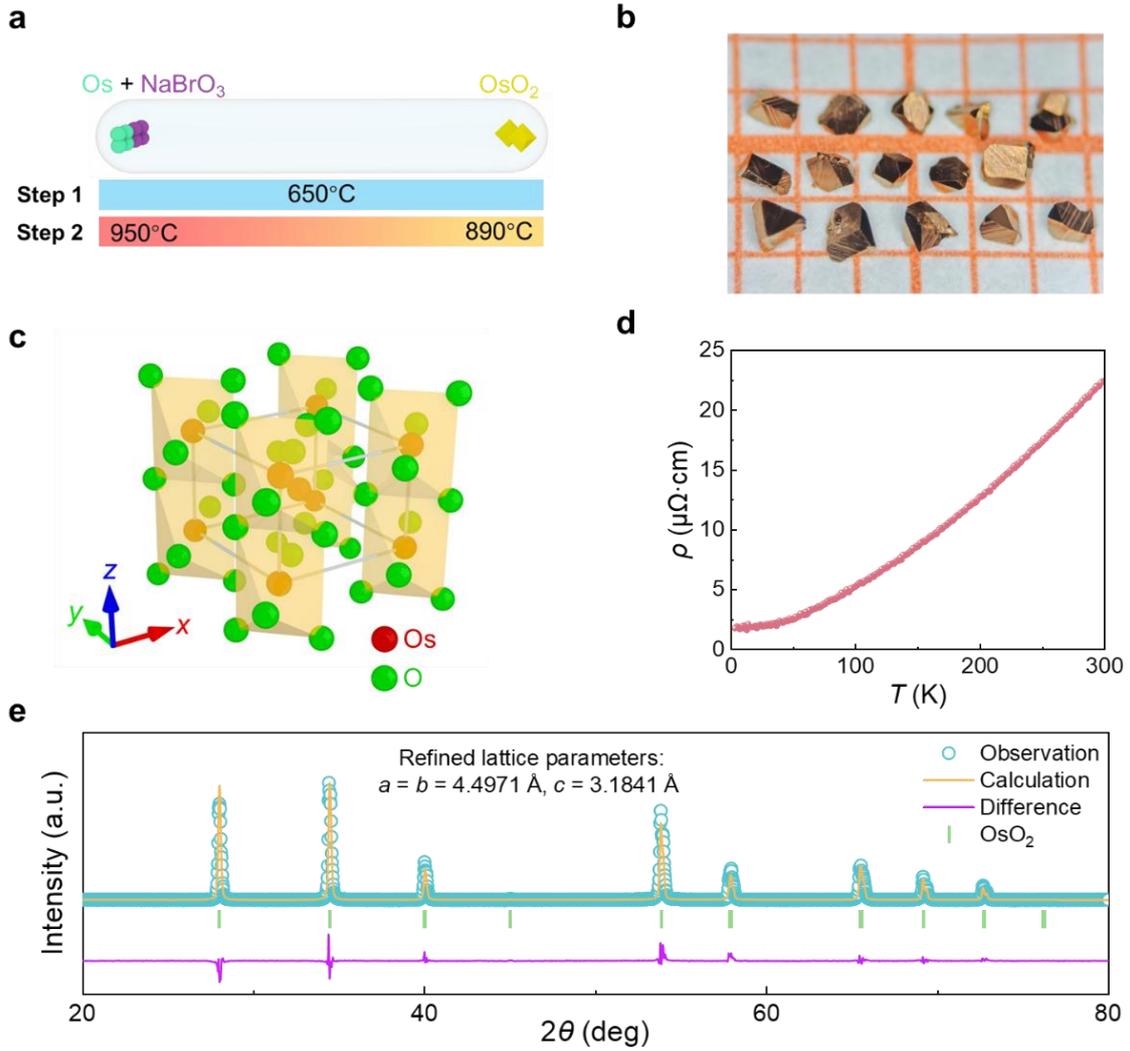

**Figure 1. Synthesis, morphology, structure, and transport properties of OsO₂ single crystals.** (**a**) Schematic of the two-step synthesis route. Os and NaBrO₃ powder are first reacted at 650 °C to form OsO₂ powder, followed by chemical vapor transport growth of single crystals under a 950 → 890 °C temperature gradient. (**b**) Optical image of OsO₂ single crystals on a millimeter-grid background. (**c**) Crystal structure of OsO₂, with Os (red) and O (green) atoms shown within a tetragonal unit cell. (**d**) Temperature-dependent electrical resistivity of OsO₂ measured from 2 to 300 K. (**e**) Powder X-ray diffraction pattern of OsO₂ obtained from ground single crystals. Experimental (blue), calculated (orange), and difference (purple) profiles are shown, with Bragg peak positions indicated in green. The data were refined by the Rietveld method, yielding lattice parameters: $a = b = 4.4971$ Å and $c = 3.1841$ Å.

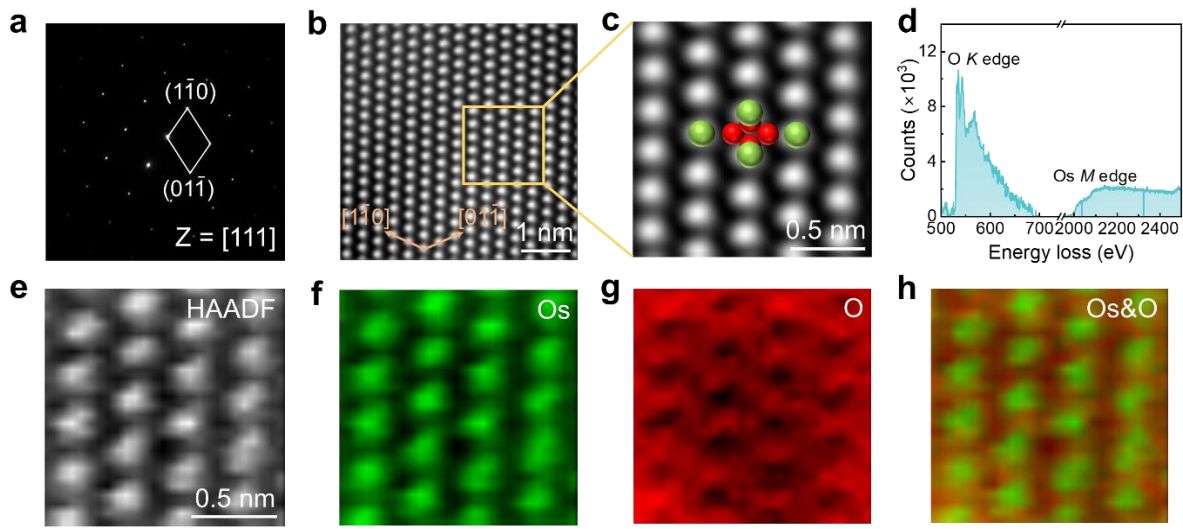

**Figure 2. Atomic-resolution imaging and elemental analysis of OsO$_2$ single crystals.** (**a**) Selected area electron diffraction pattern of an OsO$_2$ single crystal along the [111] zone axis. (**b**) High-angle annular dark-field (HAADF) scanning transmission electron microscopy image acquired along the [111] direction. (**c**) Magnified view of the boxed region in (**b**), overlaid with the projected atomic structure of OsO$_2$. (**d**) Electron energy loss spectroscopy (EELS) showing the O *K* and Os *M* edges. (**e**–**h**) EELS mapping of OsO$_2$: (**e**) HAADF image; (**f**) Os elemental map; (**g**) O elemental map; (**h**) merged elemental map of Os and O.

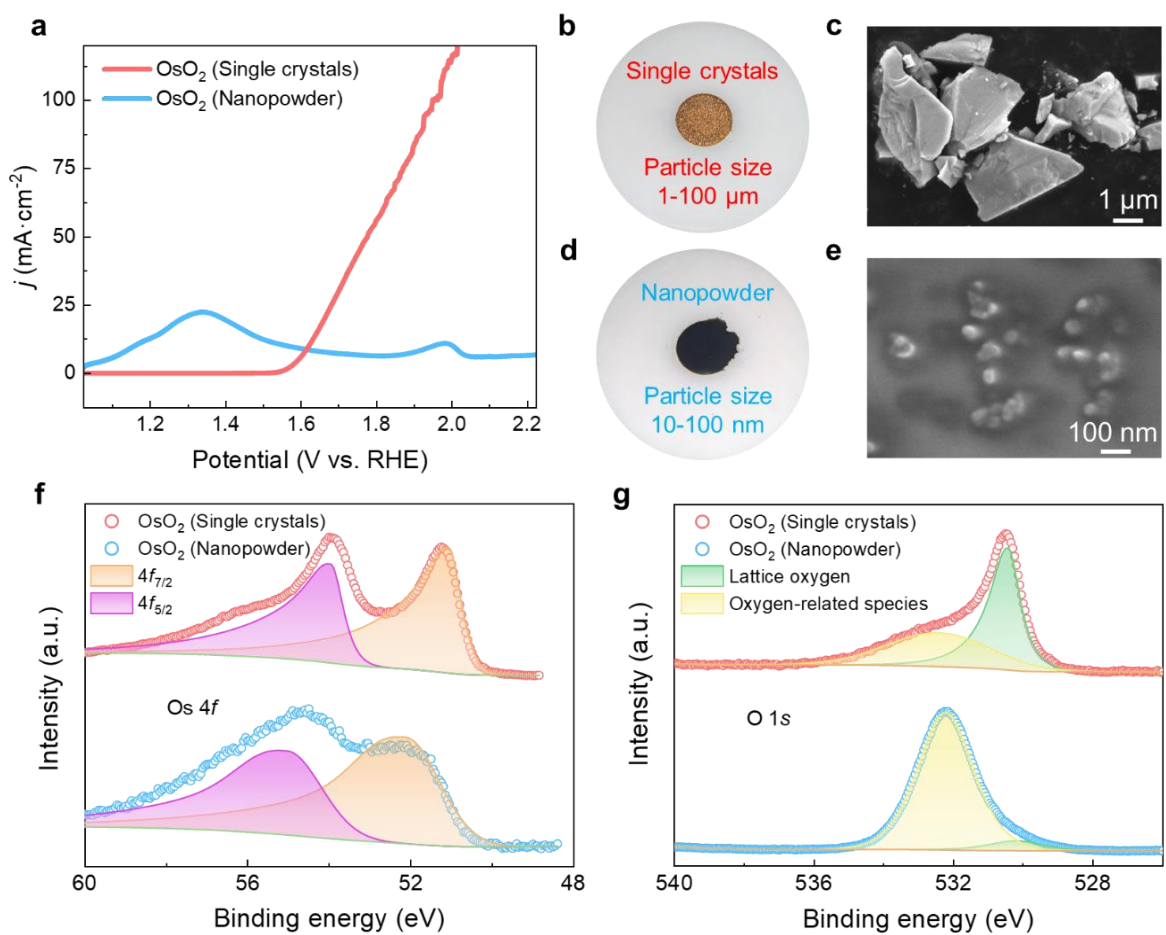

**Figure 3. Comparative electrochemical and electronic structure analysis of OsO₂ single crystals and nanopowder.** (**a**) Linear sweep voltammetry (LSV) curves of OsO$_2$ single crystals (red) and commercial nanopowder (blue) in alkaline electrolyte. (**b**, **d**) Optical images of OsO$_2$ single crystals and nanopowder, with particle sizes of 1–100 μm and 10–100 nm, respectively. (**c**, **e**) Scanning electron microscopy images of OsO$_2$ single crystals (**c**) and nanopowder (**e**), with scale bars of 1 μm and 100 nm, respectively. (**f**) X-ray photoelectron spectroscopy (XPS) of the Os 4$f$ region for single crystals (red) and nanopowder (blue), with deconvoluted 4$f_{7/2}$ (purple) and 4$f_{5/2}$ (orange) components. (**g**) XPS spectra of the O 1$s$ region, showing lattice oxygen species (green) and surface oxygen-related species (yellow).

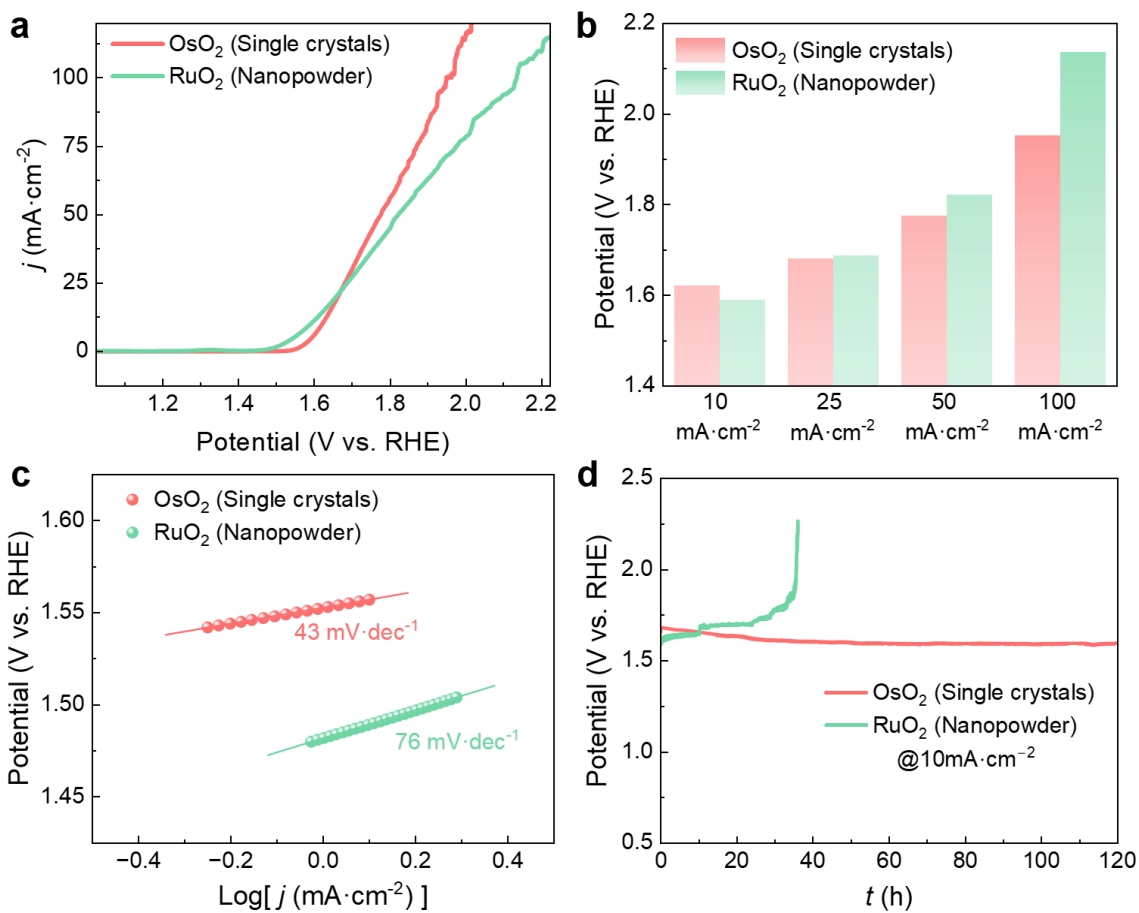

**Figure 4. Comparative oxygen evolution reaction performance of OsO$_2$ single crystals and RuO$_2$ nanopowder.** (**a**) LSV curves of OsO$_2$ single crystals and RuO$_2$ nanopowder measured in 1 mol L$^{-1}$ KOH solution. (**b**) Overpotentials at various current densities (10, 25, 50, and 100 mA·cm$^{-2}$) derived from (**a**). (**c**) Tafel plots and corresponding slopes (43 mV·dec$^{-1}$ for OsO$_2$, 76 mV·dec$^{-1}$ for RuO$_2$), extracted from (**a**). (**d**) Chronopotentiometric stability test at 10 mA·cm$^{-2}$ over 120 h.

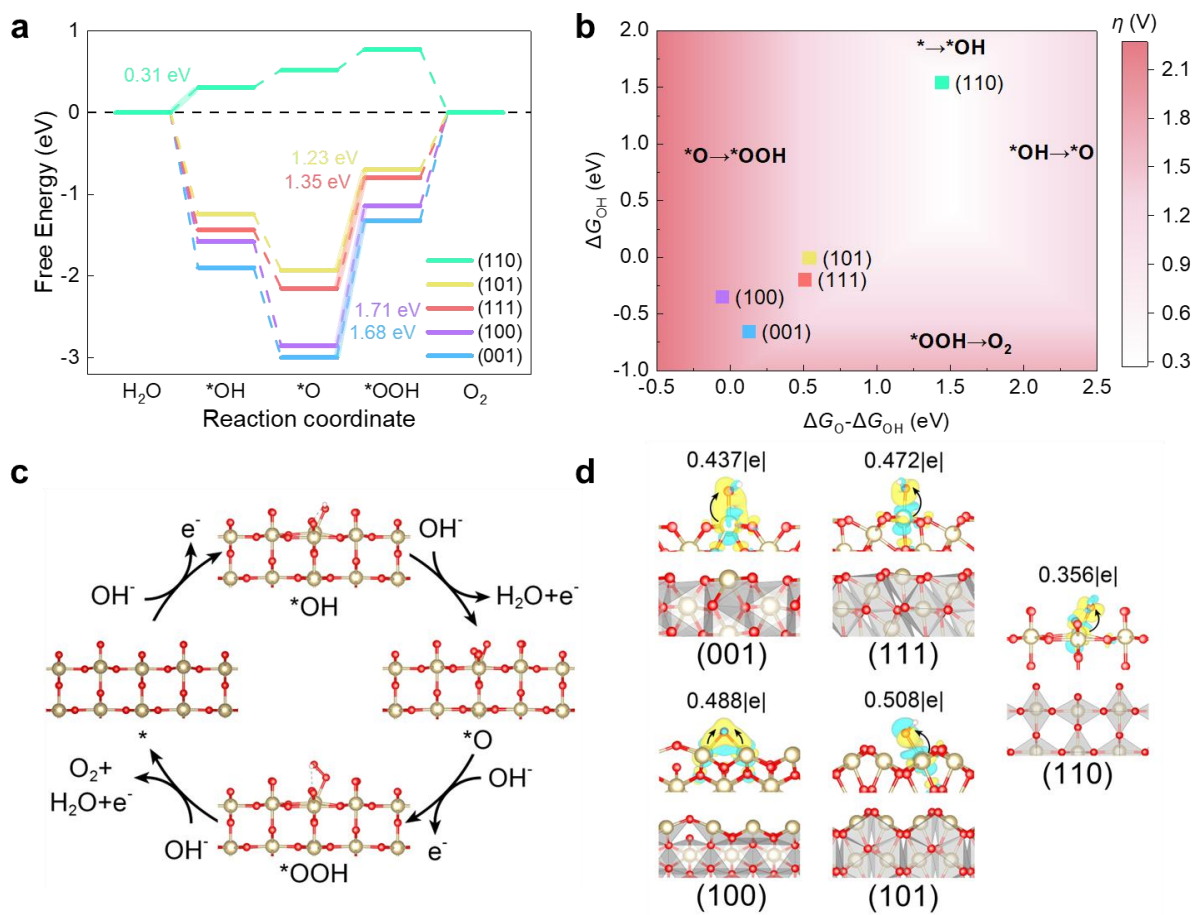

**Figure 5. DFT-calculated energetics and charge distribution of OER intermediates on OsO$_2$ surfaces.** (**a**) Gibbs free energy profiles for OER on five OsO$_2$ surfaces—(110), (101), (111), (100), and (001)—at $U$ = 1.23 V vs. RHE. (**b**) Two-dimensional volcano-type plot of overpotential ($\eta$) as a function of $\Delta G_O - \Delta G_{OH}$ and $\Delta G_{OH}$. The rate-determining step (RDS) in each region is indicated. (**c**) Four-electron OER reaction pathway on the (110) surface, showing key intermediates and elementary steps. (**d**) Charge density difference plots for adsorbed *OH intermediates on various surfaces. The values (in |e|) represent Bader charge transferred from the surface to *OH.

**Bulk OsO$_2$ Single Crystals: Superior Catalysts for Water Oxidation**

*Guojian Zhao[#], Zhihao Li[#], Ziang Meng\*, Shucheng Wang, Li Liu, Zhiyuan Duan, Xiaoning Wang, Hongyu Chen, Yuzhou He, Jingyu Li, Sixu Jiang, Xiaoyang Tan, Qinghua Zhang\*, Qianfan Zhang\*, Peixin Qin\*, Zhiqi Liu\**

**Figures S1—S13**

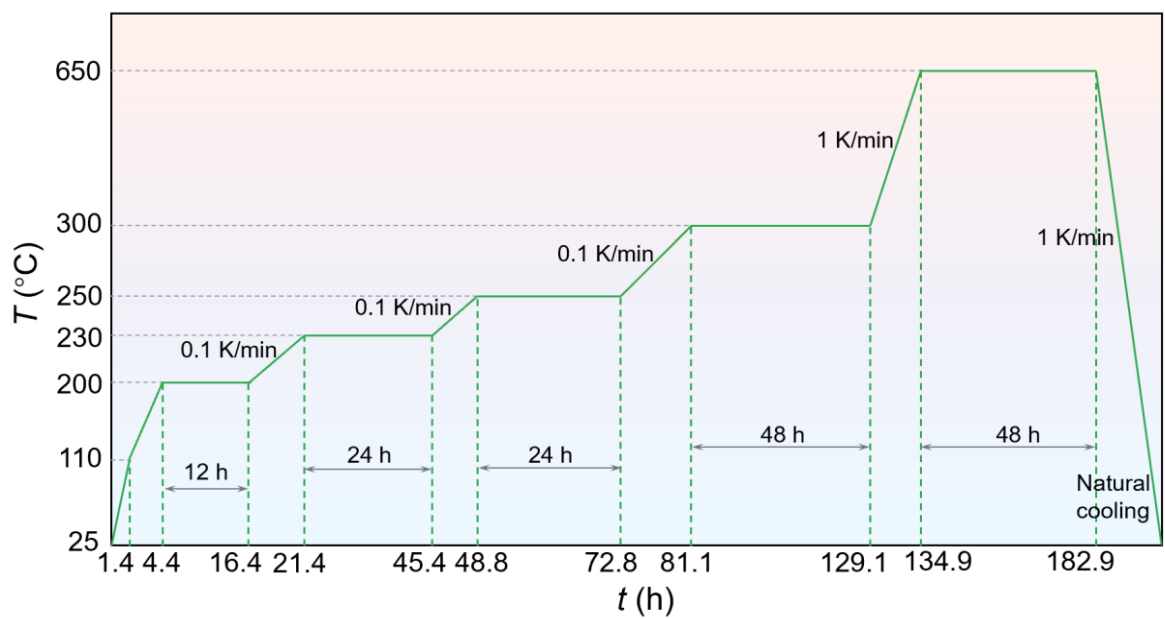

**Figure S1. Stepwise heating profile for the first-stage oxidation of Os.** The temperature–time program employs slow ramping rates and intermediate dwelling steps to allow gradual oxygen release from $NaBrO_3$ under evacuated sealed-tube conditions, enabling safe oxidation of larger Os precursor loads. This alternative protocol was not used for the samples discussed in the main text.

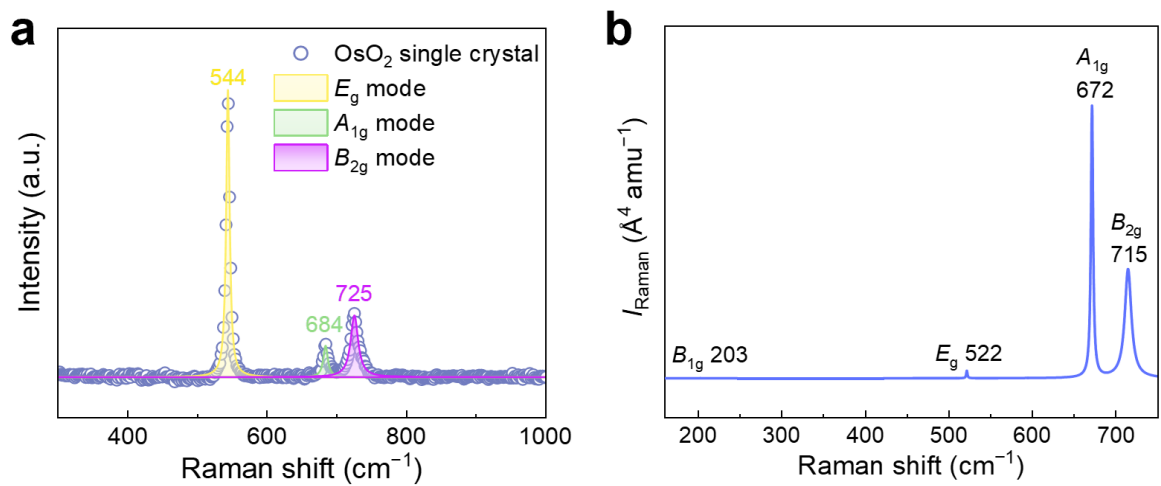

**Figure S2. Raman characterization of as-grown OsO₂ single crystals and comparison with theoretical calculations.** (a) Experimental Raman spectrum of an as-grown OsO₂ single crystal, showing three characteristic Raman-active modes of rutile-type OsO₂. The experimental spectrum is fitted using Lorentzian functions to extract peak positions. (b) Calculated Raman spectrum of rutile OsO₂ obtained from density functional theory calculations, with the corresponding vibrational modes labeled.

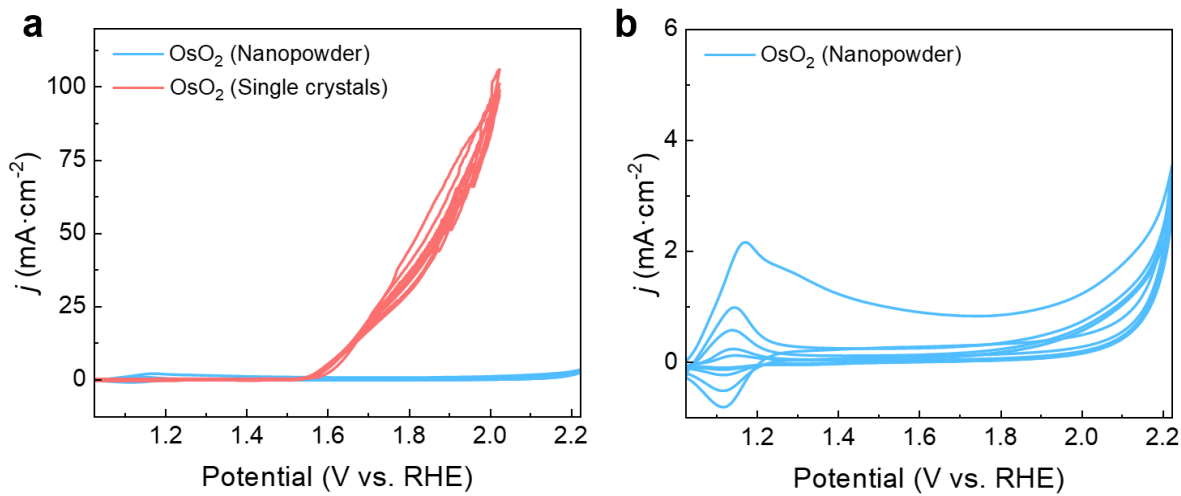

**Figure S3. Cyclic voltammetry behavior of OsO₂ single crystals and commercial OsO₂ nanopowder in alkaline electrolyte.** (**a**) Cyclic voltammetry (CV) curves of commercial OsO$_2$ nanopowder and single crystals measured in 1 mol·L$^{-1}$ KOH solution. (**b**) Magnified view of the CV curves of commercial OsO$_2$ nanopowder shown in (**a**).

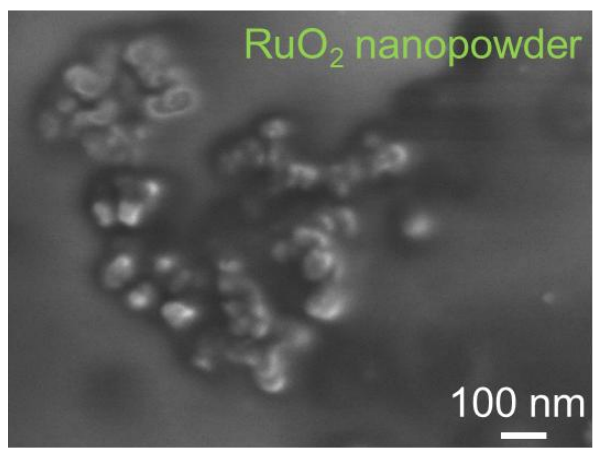

**Figure S4.** SEM image of commercial RuO$_2$ nanopowder.

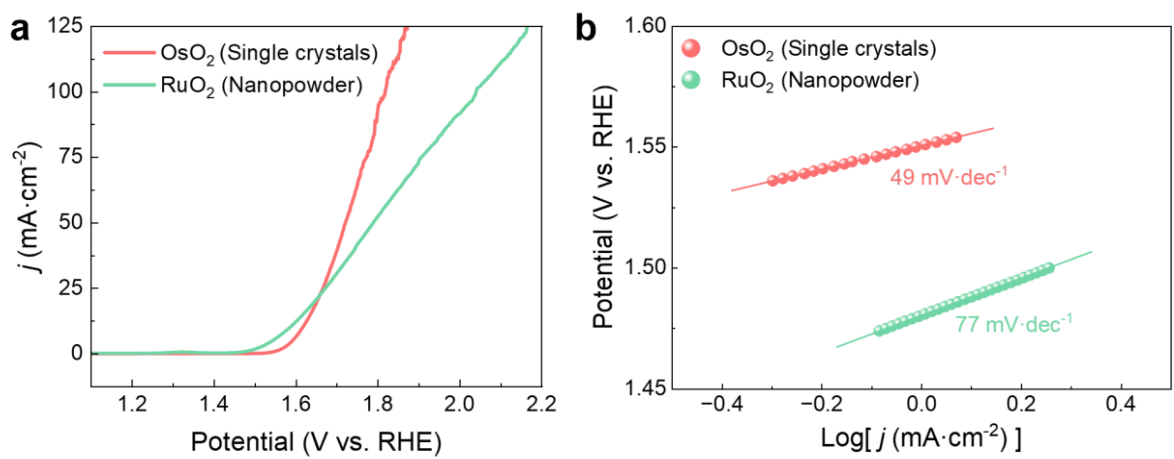

**Figure S5. Geometric OER activity comparison between OsO$_2$ single crystals and RuO$_2$ nanopowder.** (a) Linear sweep voltammetry (LSV) curves of OsO$_2$ single crystals and commercial RuO$_2$ nanopowder measured in O$_2$-saturated 1 mol·L$^{-1}$ KOH, normalized by the geometric electrode area. (b) Corresponding Tafel plots derived from the polarization curves.

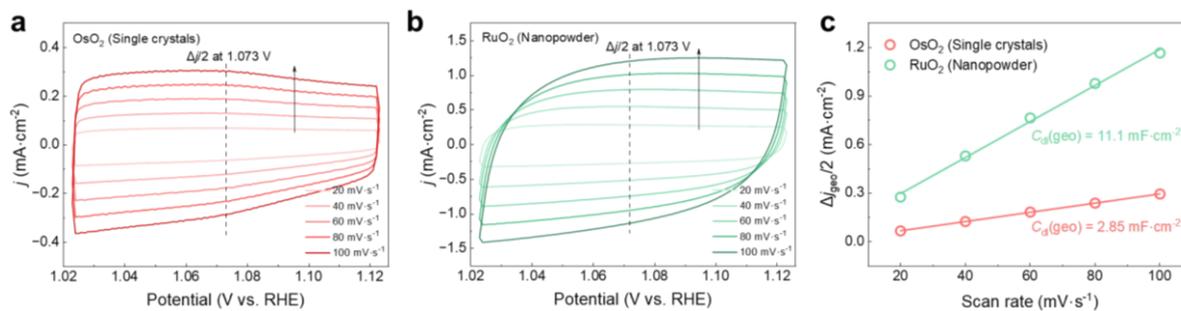

**Figure S6. Double-layer capacitance ($C_{dl}$) measurements of OsO₂ and RuO₂ electrodes.** (a) Cyclic voltammograms of OsO₂ single crystals recorded at different scan rates in a non-faradaic potential region. (b) Cyclic voltammograms of commercial RuO₂ nanopowder measured under identical conditions. (c) Linear dependence of $\Delta j/2$ on scan rate for OsO₂ single crystals and RuO₂ nanopowder, from which the areal double-layer capacitance ($C_{dl}$, geo) was extracted.

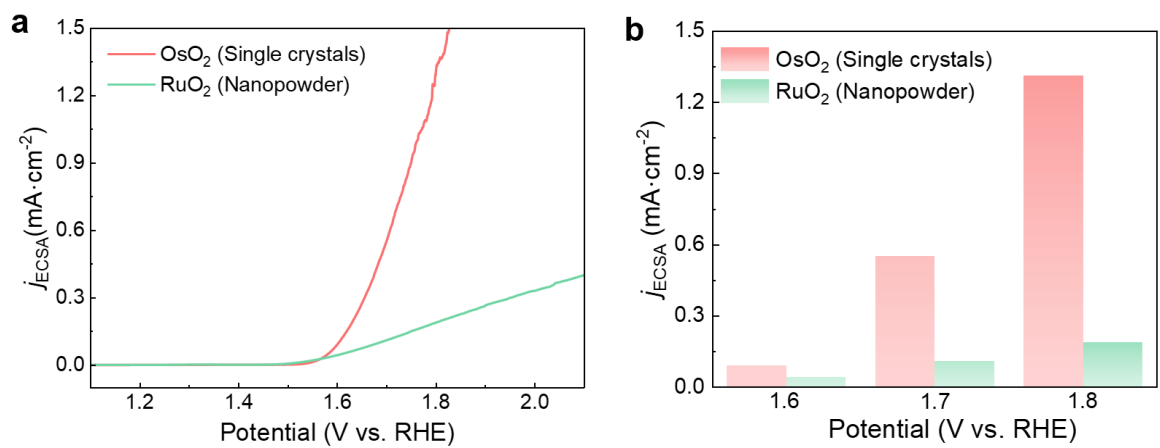

**Figure S7. ECSA-normalized OER activity of OsO₂ single crystals and RuO₂ nanopowder.** (a) ECSA-normalized OER polarization curves ($j_{ECSA}$) of OsO$_2$ single crystals and RuO$_2$ nanopowder measured in O$_2$-saturated 1 mol·L$^{-1}$ KOH. (b) Comparison of $j_{ECSA}$ at selected potentials. ECSA was estimated from $C_{dl}$ assuming $C_s$ = 40 μF·cm$^{-2}$.

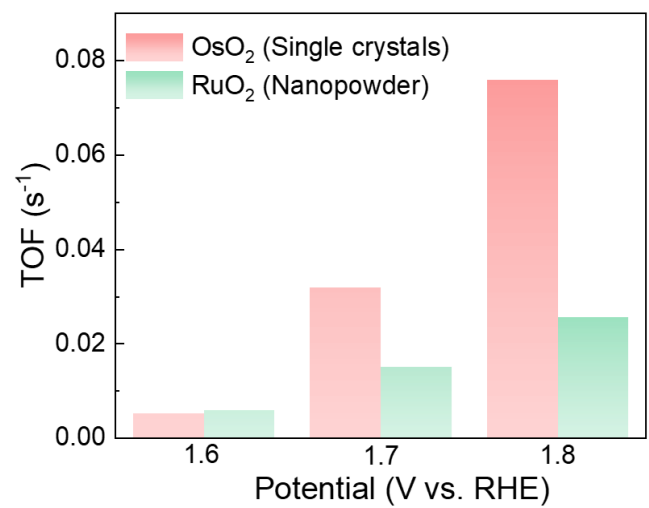

**Figure S8.** TOF values of OsO$_2$ single crystals and RuO$_2$ nanopowder at selected potentials, calculated assuming all metal atoms are active sites.

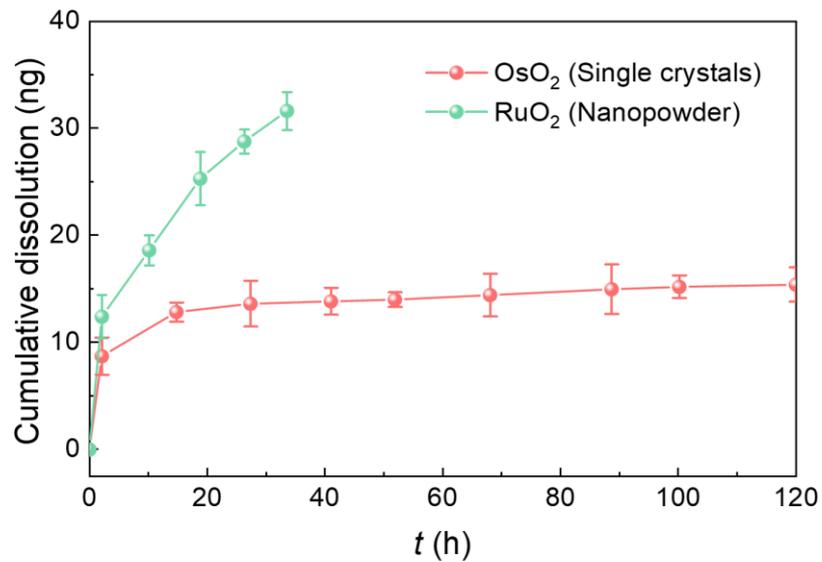

**Figure S9. Metal dissolution during OER.** Cumulative metal dissolution of Os and Ru quantified by ICP-MS as a function of OER operation time under identical conditions. Error bars represent the uncertainty estimated from three repeated measurements.

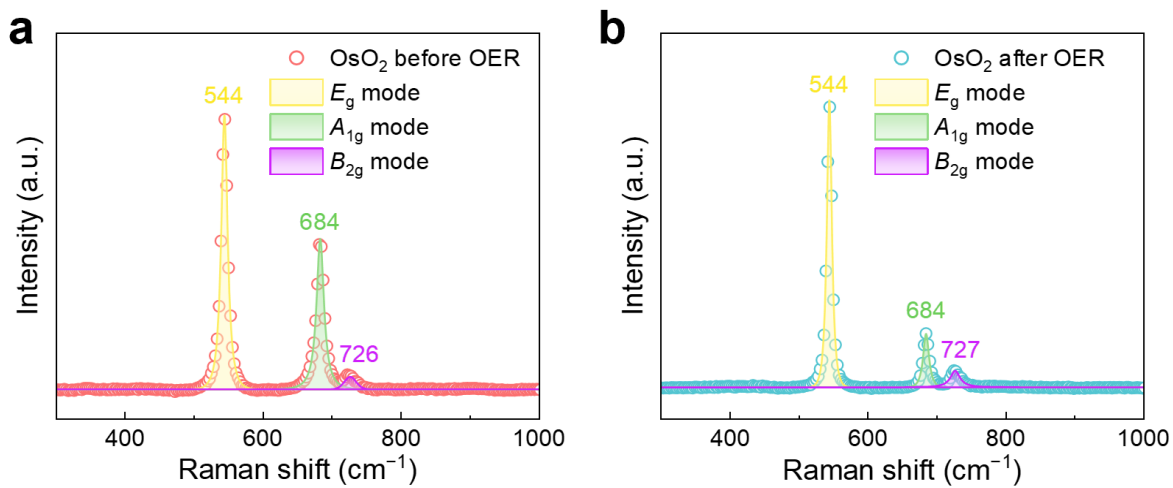

**Figure S10. Raman spectra of OsO₂ single crystals before and after OER.** (a) Raman spectrum of OsO$_2$ single crystals before OER, showing the characteristic $E_g$, $A_{1g}$, and $B_{2g}$ vibrational modes of rutile OsO$_2$. (b) Raman spectrum of OsO$_2$ single crystals after 10 h of OER operation at 10 mA·cm$^{-2}$, in which the characteristic Raman modes are preserved without the appearance of additional peaks or noticeable peak broadening.

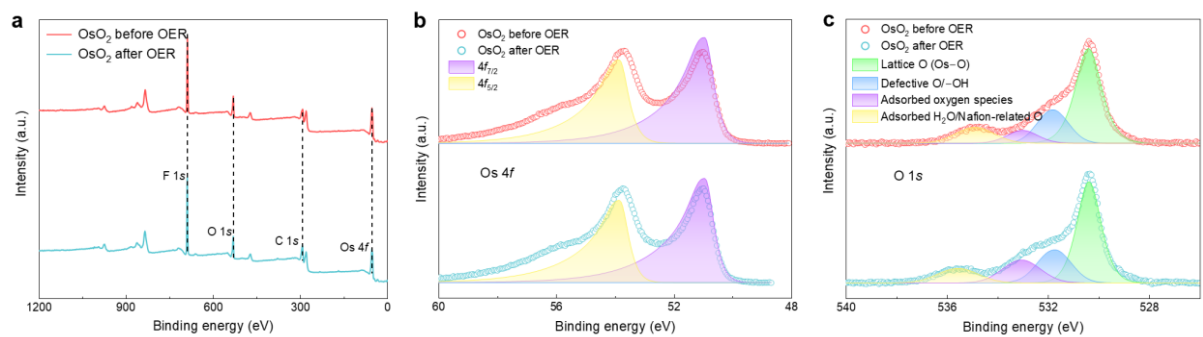

**Figure S11. XPS analysis of OsO₂ before and after OER.** (a) XPS survey spectra of OsO$_2$ electrodes before and after OER. The F 1$s$ signal observed in the XPS survey spectra originates from the Nafion binder used for electrode fabrication. (b) High-resolution Os 4$f$ spectra, showing a single Os$^{4+}$ spin–orbit doublet with negligible changes after OER. (c) High-resolution O 1$s$ spectra, deconvoluted into lattice oxygen (Os–O) and surface oxygen species. The change in relative O 1$s$ components after OER suggests a modification of surface oxygen environments.

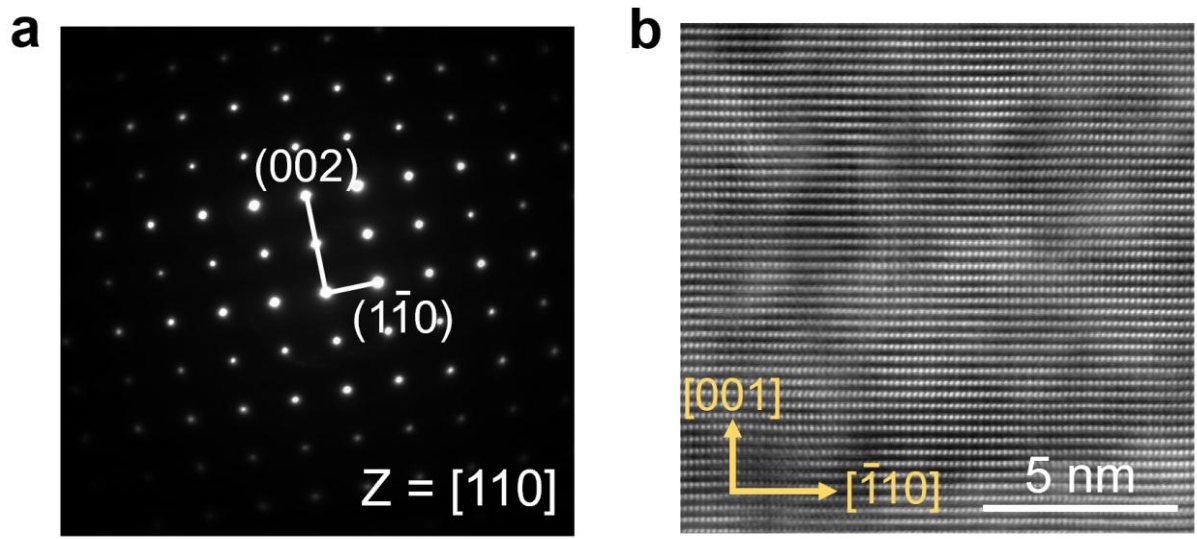

**Figure S12. TEM characterization of OsO₂ single crystals after 120 h of OER operation.**
(a) SAED pattern of $OsO_2$ (b) HAADF-STEM image of $OsO_2$.

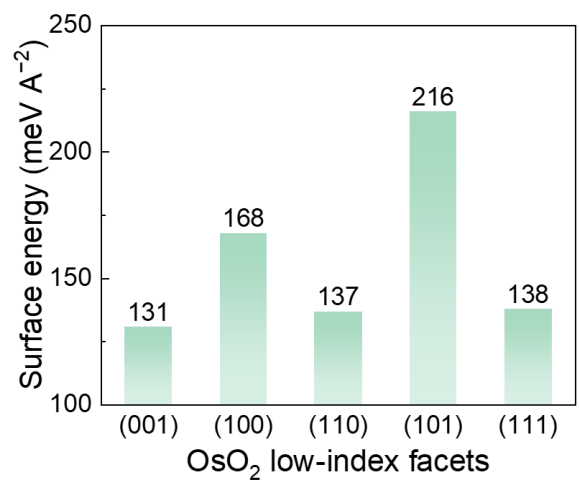

**Figure S13.** The surface energies of low-index facets of $OsO_2$